\documentclass[11pt, a4paper]{article}
\pdfoutput=1

\usepackage[height=8.85in,width=6.55in]{geometry}

\usepackage{graphicx,rotating}     
\usepackage[bookmarksopen,colorlinks=true,linkcolor=light_blue,
citecolor=darkred,urlcolor=darkred,linktoc=all]{hyperref}

\usepackage{amsmath,amssymb,tensor,subdepth}
\usepackage{mathtools}
\usepackage{slashed}
\usepackage{bm}
\usepackage{bbm}
\usepackage{cite}
\usepackage{comment}
\usepackage[utf8]{inputenc}
\usepackage{accents}
\usepackage[footnotesize]{caption}
\usepackage{cancel}
\usepackage{lettrine}
\usepackage{physics}

\usepackage[lf]{Baskervaldx} 
\usepackage[bigdelims,vvarbb]{newtxmath} 
\usepackage[cal=boondoxo]{mathalfa} 

\usepackage{stmaryrd}
\usepackage{trimclip}
\usepackage[framemethod=default]{mdframed}
\newmdenv[skipabove=7pt,
skipbelow=7pt,
rightline=false,
leftline=false,
topline=false,
bottomline=false,
backgroundcolor=gray!10,
linecolor=gray,
innerleftmargin=5pt,
innerrightmargin=5pt,
innertopmargin=5pt,
innerbottommargin=5pt,
leftmargin=0cm,
rightmargin=0cm,
linewidth=4pt]{eBox}
\newmdenv[skipabove=7pt,
skipbelow=7pt,
rightline=true,
leftline=true,
topline=true,
bottomline=true,
backgroundcolor=white,
linecolor=gray,
innerleftmargin=5pt,
innerrightmargin=5pt,
innertopmargin=5pt,
innerbottommargin=5pt,
leftmargin=0cm,
rightmargin=0cm,
linewidth=1pt]{eBox2}

\input Zallman.fd

\LettrineTextFont{\itshape}
\usepackage{chngcntr}
\counterwithin{equation}{section}

\usepackage{xcolor}
\definecolor{darkred}{rgb}{0.7, 0., 0.}
\definecolor{orangered}{rgb}{1,0.27,0.}
\definecolor{steelblue}{rgb}{0.275,0.51, 0.706}
\definecolor{forestgreen}{rgb}{0.13,0.55,0.13}
\definecolor{brightgreen}{cmyk}{0.75, 0.02, 1.00, 0.00}
\definecolor{dark_red}{rgb}{0.7, 0., 0.}
\definecolor{light_pink}{rgb}{1,0.4,0.4}
\definecolor{light_blue}{rgb}{0.284602,0.317763,0.963947}
\newcommand{\Mpl}{M_{\text{Pl}}}

\usepackage{tikz}
\usepackage{tikz-feynman}
\tikzfeynmanset{compat=1.0.0}
\pgfdeclarelayer{bg}    
\pgfsetlayers{bg,main}  
\usetikzlibrary{decorations.shapes}
\allowdisplaybreaks

\pgfdeclaredecoration{dashsoliddouble}{initial}{
  \state{initial}[width=\pgfdecoratedinputsegmentlength]{
    \pgfmathsetlengthmacro\lw{.3pt+.5\pgflinewidth}
    \begin{pgfscope}
      \pgfpathmoveto{\pgfpoint{0pt}{\lw}}%
      \pgfpathlineto{\pgfpoint{\pgfdecoratedinputsegmentlength}{\lw}}%
      \pgfmathtruncatemacro\dashnum{%
        round((\pgfdecoratedinputsegmentlength-3pt)/6pt)
      }
      \pgfmathsetmacro\dashscale{%
        \pgfdecoratedinputsegmentlength/(\dashnum*6pt + 3pt)
      }
      \pgfmathsetlengthmacro\dashunit{3pt*\dashscale}
      \pgfsetdash{{\dashunit}{\dashunit}}{0pt}
      \pgfusepath{stroke}
      \pgfsetdash{}{0pt}
      \pgfpathmoveto{\pgfpoint{0pt}{-\lw}}%
      \pgfpathlineto{\pgfpoint{\pgfdecoratedinputsegmentlength}{-\lw}}%
      \pgfusepath{stroke}
    \end{pgfscope}
  }
}

\begin{document}

\hypersetup{pageanchor=false}
\begin{titlepage}

\begin{center}

\hfill KEK-TH-2725\\ 
\hfill IPMU25-0029\\
\hfill RESCEU-13/25

\vskip 1in

{\Huge \bfseries
Dynamical dark energy\\
in the no-scale Brans--Dicke gravity\\
}
\vskip .8in

{\Large Muzi Hong$^{a,b}$, Kyohei Mukaida$^{c}$, Tsutomu T. Yanagida$^{b,d}$} \\

\vskip .3in
\begin{tabular}{ll}
$^{a}$& \!\!\!\!\!\emph{Department of Physics, Graduate School of Science, 
The University of Tokyo, Tokyo 113-0033, Japan}\\
$^{a}$&\!\!\!\!\!\emph{
RESCEU, Graduate School of Science, 
The University of Tokyo, Tokyo 113-0033, Japan}\\
$^{b}$&\!\!\!\!\!\emph{Kavli IPMU (WPI), UTIAS, The University of Tokyo, Kashiwa, Chiba 277-8583, Japan}\\
$^{c}$ & \!\!\!\!\!\emph{Theory Center, IPNS, KEK, 1-1 Oho, Tsukuba, Ibaraki 305-0801, Japan}\\
$^{c}$ & \!\!\!\!\!\emph{Graduate University for Advanced Studies (Sokendai), 
1-1 Oho, Tsukuba, Ibaraki 305-0801, Japan}\\
$^{d}$& \!\!\!\!\!\emph{Tsung-Dao Lee Institute \& School of Physics and Astronomy, Shanghai Jiao Tong University,}\\[-.3em]
& \!\!\!\!\!\emph{Pudong New Area, Shanghai 201210, China}\\
\end{tabular}

\end{center}
\vskip .6in

\begin{abstract}
\noindent
We add a new scalar field in the no-scale Brans-Dicke gravity and require it to have a global O($2$) symmetry with the original scalar field in the Brans-Dicke gravity.
This gives us a new massless scalar field in the Einstein frame due to the SO($2$) symmetry.
We then explicitly break the O($2$) symmetry to a $D_4$ symmetry, and this scalar field gains a periodic potential.
This scalar field can serve as the quintessence field to explain dark energy.
If we further add the $R^2$ term and the non-minimal coupling to the Higgs field, we can realize inflation and reheating, and this leads to a super-Planckian decay constant of the quintessence potential.
The super-Planckian decay constant is consistent with the newly released observational data according to a recent analysis.

\end{abstract}

\end{titlepage}

\renewcommand{\thepage}{\arabic{page}}
\renewcommand{\thefootnote}{$\natural$\arabic{footnote}}
\setcounter{footnote}{0}
\hypersetup{pageanchor=true}

\section{Introduction}
\label{sec:introduction}

Dark energy (DE) makes up about $70\%$ of the energy density of the current Universe, which has negative pressure and leads to the recent accelerating expansion of the Universe\cite{SupernovaSearchTeam:1998fmf,SupernovaCosmologyProject:1998vns,WMAP:2003elm,Planck:2013pxb}.
The simplest candidate of DE is a cosmological constant (c.c.) $\Lambda$ \cite{Einstein:1917ce}.
Alternatively, DE can also be explained by the energy density of a scalar field with a very flat potential, \textit{i.e.}, quintessence \cite{Fujii:1982ms,Ford:1987de,Wetterich:1987fm}.
Recently, the second data release (DR2) \cite{DESI:2025zpo,DESI:2025zgx} of the baryon acoustic oscillations measurement by the Dark Energy Spectroscopic Instrument (DESI) \cite{DESI:2013agm,DESI:2016fyo,DESI:2016igz} shows evidence for the DE changing with the Universe evolution, and thus draws attention to quintessence models.

Among the quintessence potentials, a $\cos$-type potential
\begin{equation}
    V = \mu^4 \qty[1 - \cos \qty(\frac{\phi}{f_a})]
\end{equation}
is a candidate that stands out.
Typically, axion-like particles are expected to have this kind of potential due to the $2 \pi$ periodic condition of $\phi / f_a$, where the decay constant $f_a$ is the scale of spontaneous symmetry breaking \cite{Fukugita:1994hq, Frieman:1995pm}.
The smallness of the $\phi$ particle mass in this case is due to the fact that it is a pseudo Nambu-Goldstone boson arising from spontaneous breaking of a global U$(1)$ symmetry.

The analyses of the DESI DR2 data combined with the cosmic microwave background and Type Ia supernovae observational results \cite{DESI:2025fii,Urena-Lopez:2025rad} indicate that $f_a/\Mpl \sim \,\mathcal{O}(1)$, whereas a recent analysis \cite{Lin:2025gne} claimed that $f_a$ is less constrained using a different sampling method.
It has been shown that $f_a > \Mpl$ is not excluded, \textit{i.e.}, a super-Planckian decay constant is possible.
The result of \cite{Lin:2025gne} motivates us to consider models that lead to a decay constant of $\mathcal{O}(\Mpl)$ or even larger than $\Mpl$, which strongly indicates the quintessence field is a light scalar field related to the gravity sector.

In this paper we give an example of such models in the context of \textit{no-scale Brans--Dicke gravity} \cite{Hong:2025tyi}.
Compared with the Brans--Dicke gravity \cite{Brans:1961sx}, its no-scale version satisfies the constraint from the fifth force by making the theory scale invariant in the Jordan frame, even at the quantum level\cite{Hong:2025tyi}.
In Sec.~\ref{sec:two}, we introduce a new scalar field in the no-scale Brans--Dicke gravity and require a global O($2$) symmetry between this scalar field and the original scalar field in the Brans--Dicke gravity.
We show that a new massless scalar field appears due to the SO($2$) subgroup.
An explicit breaking of the O($2$) symmetry to $D_4$ leads to a periodic potential for this scalar field, which is a candidate for quintessence with the decay constant $f_a \sim \mathcal{O}(\Mpl)$.
In Sec.~\ref{sec:R2}, we further show that a general construction of this model with the $R^2$ term and the non-minimal coupling to the Higgs field can host inflation and reheating consistent with observations, which leads to a decay constant of the quintessence field larger than $\Mpl$.
In the last section, Sec.~\ref{sec:conclusions}, we summarize the results.

\section{No-scale gravity with two scalar fields}
\label{sec:two}
To have an extra scalar field for quintessence, we add another scalar field to the no-scale Brans--Dicke gravity in the Jordan frame.
We consider a global O($2$) symmetry for these two scalar fields.
It will be shown in the following that one can avoid the unwanted fifth force in this extension.
The Lagrangian in the Jordan frame is
\begin{equation}
    \mathcal{L} = \frac{\xi}{2} \qty(\phi_1^2 + \phi_2^2)R
    + \frac{1}{2} \qty(\partial \phi_1)^2
    + \frac{1}{2} \qty(\partial \phi_2)^2
    - \lambda_1 (\phi_1^2 + \phi_2^2)^2
    + \mathcal{L}_{\rm{SM}}~,
\end{equation}
where $R$ is the Ricci scalar; $\phi_1$ and $\phi_2$ are scalar fields with the global O($2$) symmetry; and $\mathcal{L}_{\rm{SM}}$ is the Standard Model (SM) Lagrangian.
We require $\xi > 0$ so that the term with the Ricci scalar has a proper sign.
$\lambda_1$ is a free parameter at the current stage.
To satisfy the strict restriction from the fifth force (see \textit{e.g.}, \cite{Adelberger:2003zx}), we replace the dimensionful parameter in the SM, \textit{i.e.}, the mass parameter of the Higgs field, with a dimension two quantity composed by $\phi_1$ and $\phi_2$.
To respect the global O($2$) symmetry, we replace $m_H^2 \abs{H}^2$ with
\begin{equation}
    \mathcal{L}_{\rm{SM}} \supset \lambda_{m_H} (\phi_1^2 + \phi_2^2)\abs{H}^2~,
    \label{eq:massH}
\end{equation}
where $\lambda_{m_H}$ is taken to be consistent with the Higgs mass as illustrated below.
Now we perform field redefinitions as
\begin{equation}
    \phi_1 = \varphi {\rm{cos}} \theta~, \qquad
    \phi_2 = \varphi {\rm{sin}} \theta~,
    \label{eq-redef}
\end{equation}
and perform the Weyl transformation
\begin{equation}
    g_{\mu\nu}^\text{E} = \Omega^2 g_{\mu\nu}~, \qquad \Omega^2 = \frac{\xi \varphi^2}{\Mpl^2}~.
    \label{eq-Weyltrans}
\end{equation}
We obtain the Lagrangian in the Einstein frame as
\begin{equation}
    \mathcal{L}_{\rm{E}} = \frac{\Mpl^2}{2} R_{\rm{E}} + \frac{1}{2} \qty(6 + \frac{1}{\xi}) \Mpl^2 \qty(\partial {\rm{ln}} \frac{\varphi}{\Mpl})^2
    + \frac{1}{2} \frac{\Mpl^2}{\xi} \qty(\partial \theta)^2
    - \frac{\lambda_1}{\xi^2} \Mpl^4
    + \Omega^{-4} \mathcal{L}_{\rm{SM}}~.
\end{equation}
After the canonical normalization:
\begin{equation}
    \chi \equiv \sqrt{6 + \frac{1}{\xi}} \Mpl \rm{ln} \frac{\varphi}{\Mpl}~, \qquad
    \tilde{\theta} \equiv \frac{\Mpl}{\sqrt{\xi}} \theta~,
    \label{eq-cano}
\end{equation}
we end up with two massless scalar fields and a c.c. term.
With $\lambda_1 > 0$, the c.c. term has an appropriate sign.
As for the SM Lagrangian, the spinor fields and the gauge fields are Weyl invariant, while the Higgs mass term in the Einstein frame is given by
\begin{equation}
    \Omega^{-4}\mathcal{L}_\text{SM} \supset \frac{\lambda_{m_H}}{\xi} \Mpl^2 \abs{H}^2~,
\end{equation}
where we have canonically normalized the Higgs field as $H \to \Omega H$.
The Higgs field has a kinetic mixing with the $\chi$ field.
However, this does not induce the fifth force as well known.\footnote{See the discussion in \cite{Hong:2025tyi} for details.}
We have proposed to utilize the $\varphi$-dependent cutoff scale \cite{Hong:2025tyi} in the Jordan frame motivated by the scale-invariant renormalization prescription \cite{Englert:1976ep, Shaposhnikov:2008xi,Armillis:2013wya,Hamada:2016onh,Falls:2018olk}.
Then, the global Weyl symmetry is exact even at quantum level in the Jordan frame, which manifests as a shift symmetry of the $\chi$ field in the Einstein frame.\footnote{See the discussion in \cite{Hong:2025tyi}. See also \cite{Ferreira:2016kxi, Burrage:2018dvt}, in which the discussion at quantum level is not fully addressed.}
The shift symmetry of $\tilde{\theta}$ is due to the SO($2$) subgroup of the O($2$) symmetry.

Now we introduce small explicit breakings of the global O($2$) symmetry.
We require that a $D_4$ finite subgroup of O($2$) which includes, \textit{e.g.}, the exchanging symmetry and parity symmetry of $\phi_1$ and $\phi_2$, remains after the explicit symmetry breaking, to generate a desired quintessence potential.
The Lagrangian is given by
\begin{equation}
    \mathcal{L} = \frac{\xi}{2} \qty(\phi_1^2 + \phi_2^2)R
    + \frac{1}{2} \qty(\partial \phi_1)^2
    + \frac{1}{2} \qty(\partial \phi_2)^2
    - \lambda_1 (\phi_1^2 + \phi_2^2)^2
    - \lambda_2 \phi_1^2 \phi_2^2
    + \mathcal{L}_{\rm{SM}}~.
    \label{eq-brkL}
\end{equation}
We assume $\lambda_2 > 0$ and $\lambda_1$ with an arbitrary sign for the convenience of later discussion. 
The mass term of the Higgs field is still (\ref{eq:massH}) as $\phi_1 \phi_2 |H|^2$ violates the $D_4$ symmetry.
We perform Weyl transformation (\ref{eq-Weyltrans}) and obtain the action in the Einstein frame as
\begin{equation}
    \mathcal{L}_{\rm{E}} = \frac{\Mpl^2}{2} R_{\rm{E}} + \frac{1}{2} \qty(6 + \frac{1}{\xi}) \Mpl^2 \qty(\partial {\rm{ln}} \frac{\varphi}{\Mpl})^2
    + \frac{1}{2} \frac{\Mpl^2}{\xi} \qty(\partial \theta)^2
    - \frac{\lambda_1}{\xi^2} \Mpl^4
    - \frac{\lambda_2}{\xi^2} \Mpl^4 {\rm{cos}}^2\theta {\rm{sin}}^2\theta
    + \Omega^{-4} \mathcal{L}_{\rm{SM}}~.
\end{equation}
We again canonically normalize the fields as in (\ref{eq-cano}), and thus we have
\begin{equation}
    \mathcal{L}_{\rm{E}} = \frac{\Mpl^2}{2} R_{\rm{E}}
    + \frac{1}{2} \qty(\partial \chi)^2
    + \frac{1}{2} \qty(\partial \tilde{\theta})^2
    - \frac{\lambda_1}{\xi^2} \Mpl^4
    - \frac{\lambda_2}{8 \xi^2} \Mpl^4 \qty[1 - {\rm{cos}} \qty(\frac{4 \sqrt{\xi}}{\Mpl} \tilde{\theta})]
    + \Omega^{-4} \mathcal{L}_{\rm{SM}}~.
    \label{eq-quintessence}
\end{equation}
Remarkably, we have an axion-type potential for quintessence with the decay constant $f_a = \Mpl / (4 \sqrt{\xi})$.
If $\lambda_2 \gg \abs{\lambda_1}$, the DE is dominated by the quintessence potential.
Note that the smallness of $\lambda_2$ is protected by the enhancement of the shift symmetry of $\tilde{\theta}$, \textit{i.e.}, the SO($2$) symmetry.

\section{Adding $R^2$ and non-minimal coupling to Higgs}
\label{sec:R2}
Now we consider a general Lagrangian without dimensionful parameters while maintaining the symmetry of $\phi_1$ and $\phi_2$ proposed in the last section.
This implies the introduction of the $R^2$ term and the non-minimal coupling of the Higgs field.
We will show that a successful inflation requires a small $\xi$, naturally leading to the super-Planckian decay constant of the quintessence field.

Let us add them to (\ref{eq-brkL}) \cite{Hong:2025tyi}:
\begin{align}
    \mathcal{L} = & \qty[\frac{\xi}{2} \qty(\phi_1^2 + \phi_2^2) + \xi_H \abs{H}^2]R
    + \frac{1}{2} \qty(\partial \phi_1)^2
    + \frac{1}{2} \qty(\partial \phi_2)^2
    + \abs{\partial H}^2
    + \alpha R^2\\
    &- \lambda_1 (\phi_1^2 + \phi_2^2)^2
    - \lambda_2 \phi_1^2 \phi_2^2
    - \frac{\lambda_H}{4} \qty[ \abs{H}^2 - \lambda_v^2 \qty(\phi_1^2 + \phi_2^2) ]^2
    + \mathcal{L}'_{\rm{SM}}~,
\end{align}
where $H$ is the Higgs doublet and $\mathcal{L}'_{\rm{SM}}$ denotes the SM Lagrangian without the kinetic term and the potential of the Higgs field.
$\lambda_H$ is the Higgs four-point coupling and $\lambda_v$ is taken to be consistent with the Higgs mass.
We require $\alpha > 0$.
As will be shown shortly, this is related to the proper sign of the scalaron potential. 
$\xi_H$ is a free parameter at the current stage.
By introducing an auxiliary field $X$, we rewrite the $R^2$ term as
\begin{equation}
    \alpha R^2 \to 2 \alpha X R - \alpha X^2~.
\end{equation}
Again we use the field redefinition of $\phi_1$ and $\phi_2$ in (\ref{eq-redef}), and perform a Weyl transformation with
\begin{equation}
    \Omega^2 = \frac{\xi (\phi_1^2 + \phi_2^2) + \xi_H h^2 + 4 \alpha X}{\Mpl^2}~,
\end{equation}
where $h \equiv (2\abs{H}^2)^{1/2}$ is the radial mode of the Higgs field.
To diagonalize the kinetic terms we further perform field redefinitions as \cite{Casas:2017wjh,Hong:2025tyi}
\begin{equation}
    \frac{\phi_1^2}{\Mpl^2} = \qty( 1 - \Theta^{-1} ) \cos^2 \Phi \cos^2 \theta\, e^{ \sqrt{\frac{2}{3}} \frac{\chi}{\Mpl} }, \qquad
    \frac{\phi_2^2}{\Mpl^2} = \qty( 1 - \Theta^{-1} ) \cos^2 \Phi \sin^2 \theta\, e^{ \sqrt{\frac{2}{3}} \frac{\chi}{\Mpl} },
\end{equation}
\begin{equation}
        \Omega^2 = \frac{1}{6\Theta} e^{ \sqrt{\frac{2}{3}} \frac{\chi}{\Mpl} }, \qquad
        \frac{h^2}{\Mpl^2} = \qty( 1 - \Theta^{-1} ) \sin^2 \Phi\, e^{ \sqrt{\frac{2}{3}} \frac{\chi}{\Mpl} }.
\end{equation}
These definitions imply
\begin{equation}
    \Theta = \frac{\qty( \xi + 1/6 ) \varphi^2 + \qty( \xi_H + 1/6 ) h^2 + 4 \alpha X}{\xi \varphi^2 + \xi_H h^2 + 4 \alpha X}~,
\end{equation}
thus $\Theta \geq 1$. 

The Lagrangian in the Einstein frame is now
\begin{align}
    \mathcal{L}_\text{E} = &\frac{\Mpl^2}{2} R_\text{E} + \frac{1}{2} \Theta\, \qty(\partial \chi)^2 + \frac{3}{4} \Mpl^2\, \frac{1}{\Theta \qty( \Theta - 1 )}\, \qty(\partial \Theta)^2
    + \frac{1}{2} \cdot 6 \Mpl^2 \qty( \Theta - 1 ) \qty[ \qty( \partial \Phi )^2 + \sin^2 \Phi\, \qty( \partial \bm{\pi} )^2 ] \\
    & 
    \label{eq-ThetaPhiV}
    - \frac{\Mpl^4}{16 \alpha} \qty{ 1 - 6 \qty( \Theta - 1 ) \qty[ \xi + \qty( \xi_H - \xi ) \sin^2 \Phi ] }^2
    - \frac{9}{4} \lambda_H \Mpl^4 \qty( \Theta - 1 )^2 \sin^4 \Phi \\
    &
    \label{eq-thetaL}
    + \frac{1}{2} \cdot 6 \Mpl^2 \qty( \Theta - 1 ) \cos^2 \Phi \qty(\partial \theta)^2
    + \Omega^{-4} \mathcal{L}_{\text{SM}'}~,
\end{align}
where $\bm{\pi}$ is the angular modes of the Higgs field restricted to a unit sphere, \textit{i.e.}, $\bm{\pi} \cdot \bm{\pi} = 1$.
We dropped $\lambda_v$, $\lambda_1$ and $\lambda_2$ since they are negligible for the inflation dynamics.
The $\Theta$ field gives rise to a successful inflation \cite{Ema:2017rqn, He:2018gyf} as we see shortly below.

The kinetic term of $\Theta$ has a pole at $\Theta = 1$ and near it the potential of the canonically normalized inflaton becomes flat.
The potential has a minimum at $\Theta_0 \simeq 1 + 1/(6 \xi)$, therefore a small $\xi$ allows the inflaton have enough field excursion range along the plateau of the potential and thus leads to a successful inflation.
As for the precise value of $\xi$, it has been shown in the previous work \cite{Hong:2025tyi} that the $\Theta$ field canonically normalized as
\begin{equation}
    \Theta = \cosh^2 \qty( \sqrt{ \frac{2}{3} } \frac{\sigma}{2\Mpl} )
\end{equation}
can serve as inflaton with $\xi \lesssim 10^{-3}$.
If further restricting the discussion to inflation driven by the $R^2$ term
\footnote{
    This is easily realized by taking $\xi_H < 0$ for instance, which is also motivated by stabilizing the Higgs potential during inflation. See Ref.~\cite{Hong:2025tyi} for more details.
},
it has been shown that it is possible to realize reheating without overproduction of dark radiation composed of $\chi$ and $\theta$ particles \cite{Hong:2025tyi}.\footnote{The dark radiation is twice the amount of that discussed in \cite{Hong:2025tyi} due to the contribution of the new scalar field.}

During the period when the discussion of quintessence is relevant, the $R^2$ term and the Higgs field becomes irrelevant.
We have a canonically normalized quintessence field with a c.c.\ term and a periodic potential as in (\ref{eq-quintessence}).
We can see that the quintessence field has a super-Planckian decay constant with $\xi \lesssim 10^{-3}$, which implies
\begin{equation}
    f_a \gtrsim 10 \Mpl ~.
\end{equation}
As illustrated above, a consistent inflation and reheating is possible in this case.

\section{Conclusions}
\label{sec:conclusions}
Recent analysis \cite{Lin:2025gne} of the newly released DESI DR2 observational data indicates that a super-Planckian decay constant in the potential of the axion-type quintessence is allowed.
The result in \cite{Lin:2025gne} motivates us to extend the original no-scale Brans-Dicke model \cite{Hong:2025tyi} and build a quintessence model which naturally has a super-Planckian decay constant.

We first add another scalar field in the no-scale Brans-Dicke gravity, and we require the theory to have a global O($2$) symmetry between the new scalar field and the original scalar field in the Brans-Dicke gravity.
This gives us two massless scalar fields in the Einstein frame.
Then we introduce a small explicit breaking of the O($2$) symmetry while keeping the $D_4$ subgroup to generate a small mass term for one of the scalar fields.
The breaking global SO($2$) symmetry with the remaining $D_4$ leads to the appearance of its periodic potential.
The general construction of this model involves the $R^2$ term and non-minimal coupling to Higgs, which then allows inflation and reheating.
A successful inflation consistent with the observations requires a small $\xi$ and thereby a super-Planckian decay constant of the quintessence field, \textit{i.e.,} $f_a \gtrsim 10 \Mpl$, which can explain dynamical DE.
We hope that the more precise measurement of the DE evolution in future observations will confirm the large decay constant as this model predicts.

One can further extend this model by adding a new scalar field in the Jordan frame and require an SO($3$) symmetry between $\phi_1$, $\phi_2$ and this new scalar field, since it is natural to embed the O($2$) symmetry discussed in this paper in an SO($3$) symmetry.
By explicitly breaking the SO($3$) symmetry, one may manage to have two light scalar fields coupling with each other in the Einstein frame, arising from the two angular coordinates.
One of them corresponds to the quintessence field discussed in this paper, and it will be interesting to explore the role of the other one.
This model leads to possible scenarios such as two quintessence fields, or a light Dark Matter interacting with the DE.

\section*{Acknowledgment}
M.\,H.~is supported by Grant-in-Aid for JSPS Fellows 23KJ0697.
K.\,M.~is supported by JSPS KAKENHI Grant No.~JP22K14044.
T.\,T.\,Y.~is supported by the Natural Science Foundation of China (NSFC) under Grant No.~12175134, MEXT KAKENHI Grants No.~JP24H02244, and
World Premier International Research Center Initiative
(WPI Initiative), MEXT, Japan.

\small
\bibliographystyle{utphys}
\bibliography{ref}

\end{document}